\documentclass[12pt,a4paper,final]{iopart}

\usepackage{iopams}
\usepackage{graphicx,caption,cite}% Include figure files
\usepackage[colorlinks,allcolors=blue]{hyperref}
\usepackage{amssymb}

\begin{document}

\title{Fractional Fokker-Planck equation from non-singular kernel operators}

\author{M. A. F. dos Santos$^{1}$}
\address{$^1$Instituto de F\'isica, Universidade Federal do Rio Grande do Sul, Caixa Postal 15051, CEP 91501-970, Porto Alegre, RS, Brazil}
\ead{santos.maike@ufrgs.br}

\author{Ignacio S. Gomez$^{2}$}
\address{$^2$Instituto de F\'isica, Universidade Federal da Bahia,
			 R.\ Bar\~ao de Jeremoabo s/n, 40170-115 Salvador, Bahia, Brazil}
\ead{nachosky@fisica.unlp.edu.ar}
\vspace{10pt}
\begin{indented}
\item[] \today
\end{indented}

\begin{abstract}
Fractional diffusion equations imply non-Gaussian distributions that generalise the standard diffusive process. Recent advances in fractional calculus lead to a class of new fractional operators defined by non-singular memory kernels, differently from the fractional operator defined in the literature. In this work we propose a generalisation of the Fokker-Planck equation in terms of a non-singular fractional temporal
operator and considering a non-constant diffusion coefficient.
We obtain analytical
solutions for the Caputo-Fabrizio and the Atangana-Baleanu fractional kernel operators, from which non-Gaussian distributions emerge having a long and short tails. In addition, we show that these non-Gaussian distributions are unimodal or bimodal
according if the diffusion index $\nu$ is positive or negative respectively, where a diffusion coefficient of the power law type $\mathcal{D}(x)=\mathcal{D}_0|x|^{\nu}$ is considered.
Thereby, a class of anomalous diffusion phenomena connected with fractional derivatives and with a diffusion coefficient of the power law type
%$\mathcal{D}(x)=\mathcal{D}_0|x|^{\alpha}$,
is presented. The techniques employed in this work open new possibilities for studying memory effects in diffusive contexts.
\end{abstract}

\vspace{2pc}
\noindent{\it Keywords}:
Fractional FPE, anomalous diffusion, Caputo-Fabrizio operator, Atangana-Baleanu operator
%\maketitle

\section{\label{sec:level1}Introduction}

Gaussian forms are the most studied probability distributions in statistical physics.
Based on a thermal framework, Einstein proposed
the standard diffusion equation and thus he determined the Gaussian forms
that characterise the movement of Brownian particles immersed in an equilibrium liquid \cite{einstein1905molekularkinetischen}. These investigations were also continued
during the period 1908-1914 by Perrin and Nordlund
\cite{perrin1908agitation,perrin1909mouvement,nordlund1914new}, while alternative formulations were proposed by Sutherland, Smoluchowski, and Langevin \cite{sutherland1905lxxv,von1906kinetischen,langevin1908theorie}.
The Brownian motion is characterised by the linear evolution of the mean square displacement (MSD), that is the so-called linear diffusion i.e. $\langle (\Delta x)^2 \rangle \sim t$, and the break of this linearity
defines the
anomalous diffusion (or non-Fickian diffusion) \cite{bouchaud1990anomalous}.
Anomalous diffusion process is characterised by a
power law dependence of the MSD, i.e. $\langle (\Delta x)^2 \rangle \sim t^{\mu}$, for which $0<\mu<1$ and $1<\mu<2$ determine the subdiffusive and superdiffusive regimes respectively.
In some particular cases, ballistic and hyper diffusive processes can occur for
the values $\mu=2$ and
$2<\mu$. Several works show that the anomalous diffusion is strongly connected with non-Gaussian distributions \cite{w1,w2,w3,w4,w5}
that can be also obtained
as solutions of generalisations of the diffusion equation
\cite{bologna2000anomalous,chechkin2017brownian}.

Non-Gaussian distributions are associated intimately with generalisations of the Fokker-Planck equation (FPE) and they
allow to
model multiple and complex phenomena as non-ergodicity, non-locality, memory effects and long range interactions.
Complementary, nonlinear, heterogeneous and fractional approaches also provide generalisations of the FPE.
Nonlinear diffusion equations, whose
solutions exhibit
a non-Gaussian behaviour, has been associated to nonlinear FPE in several scenarios: quantum walks \cite{nonlinear3}, non-extensive statistical mechanics \cite{nonlinear1,nonlinear4,dos2017entropic,nonlinear2,sicuro2016nonlinear},  generalisations of the Central Limit Theorem which imply
non-Gaussian forms (called as $q$-Gaussian distributions) \cite{umarov2007multivariate,umarov2008q,jauregui2015convergence,jauregui2010new}, among others.
In addition,
the study of the FPE with a non-constant diffusion coefficient has allowed to introduce a natural generalisation of the Brownian motion in an heterogeneous medium \cite{risken1996fokker,cherstvy2013anomalous}. This approach was reported in: comb model structures \cite{sandev2017heterogeneous}, reaction--diffusion model \cite{hormuth2017mechanically}, non-ergodicity, criticality of diffusive systems \cite{metzler2014anomalousd} and  intracellular transport \cite{lanoiselee2018model}, etc.

Other way of investigating the Brownian motion in a heterogeneous medium can be achieved by means of the Ito-Langevin equation that links
the type of displacement of the trajectories
with a heterogeneous FPE \cite{risken1996fokker,borland1998microscopic}.
This formalism characterises the diffusion equation through fractional operators, heading to a fractional dynamic. The use of fractional derivatives for describing diffusion equations has been turned out a useful method to analyze non-local phenomena and memory effects
in several systems \cite{li2018time,ervin2018regularity,fractional1,fractional2,fractional3,dos2017anomalous,klages2008anomalous,SUN20094586,yang2017new}.
In this context, Caputo and Fabrizio have recently proposed a non-singular fractional derivative expressed by an exponential kernel \cite{caputo2015new}. Subsequently, Atangana \textit{et. al} also have defined a non-singular kernel in terms by the Mitag-Leffler function \cite{atangana2016new}. These non-singular operators present properties that allow to describe transport phenomena, anomalous diffusion process and dynamical models \cite{gomez2017space,atangana2016chaos,xiao2017general,dokuyucu2018cancer,hristov2017derivation,alkahtani2017generalized,gomez2018analytical} simultaneously.

The main goal of this paper
to propose a generalisation of the FPE,
in terms of a non-singular temporal kernel operator
and a diffusion coefficient of the power law type, that contains all known anomalous (sub-super-ballistic-hyper) diffusive
regimes as special cases.
For accomplish this, we use the Caputo-Fabrizio and the Atangana-Baleanu fractional kernel operators and we obtain non-Gaussian analytical
solutions with unimodal and bimodal characteristics.
Thus, our contribution consist of providing
a unified framework for studying diffusion processes generated by non-singular fractional operators.
The work is structured as follows. In Section \ref{sec2} we review some concepts of the fractional calculus related to the Caputo-Fabrizio and Atangana-Baleanu fractional derivatives along with a generalisation of the FPE within the fractional calculus framework.
Section \ref{sec3} is devoted to obtain the solutions of the generalised FPE from the non-singular temporal kernels of the Caputo-Fabrizio and Atangana-Baleanu fractional derivatives,
and considering a arbitrary variable diffusion coefficient, i.e. $\mathcal{D}(x)$.
We obtain analytical solutions for the Caputo-Fabrizio and Atangana-Baleanu fractional operators.
In Section \ref{sec4} we characterise the type of the diffusion obtained from the calculus of the MSD.
As a consequence, subdiffusion, superdiffusion and hyperdiffusion are obtained as special cases. In Section \ref{sec5} some conclusions and perspectives are outlined.

\section{
Preliminaries}
\label{sec2}
We review some notions and concepts
used
throughout the paper.

\subsection{Fractional calculus}\label{subsec2.1}

Fractional calculus considers the different possibilities of extending powers of the usual
differentiation to the field of the real numbers or complex ones.
The Riemann-Liouville integral constitutes one of the most traditional forms of fractional integration, expressed by
\begin{eqnarray}\label{fractional-derivative-RL}
{}_{a} \mathcal{D}_{t}^{-\alpha} f(t)=
\frac{1}{\Gamma(\alpha)}\int_{a}^{t}(t-s)^{\alpha-1}f(s)ds
\end{eqnarray}
where $f(t)$ is an arbitrary function and $\Gamma(\ldots)$ denotes the Gamma function. By the Laurent series,
from
(\ref{fractional-derivative-RL}) we can see that for $\alpha=0$ and $\alpha=1$ the original function $f(t)$ and its primitive $\int_a^t f(s)ds$ are recovered respectively. In this sense, the fractional derivative of Riemann--Liouville is defined by $\frac{d}{dt} {}_{a} \mathcal{D}_{t}^{-(1-\alpha)} f(t)$, in which $0<\alpha<1$. Other fundamentals  forms to fractional derivative as Gr\"unwald--Letnikov, Caputo, Riesz and other are detailed in Ref. \cite{podlubny1998fractional}.
Recent generalisations of fractional derivatives have been developed \cite{caputo2015new,atangana2016new}. In this work we focus on two types: the Caputo-Fabrizio and the Atangana-Baleanu fractional derivatives. The former does not need
to define fractional order initial conditions as in the
Riemann-Liouville case, while the later uses
a general Mittag-Leffler function as a kernel.
These type of fractional derivatives can be expressed in the compact form
\begin{eqnarray}\label{fractional-derivative-kernel}
\mathcal{O}_t^{\mathcal{K}} f(t)=
\int_{0}^{t}dt^{\prime} \mathcal{K}(t-t^{\prime})\frac{d\ }{dt^{\prime}} f(t^{\prime})
\end{eqnarray}
where $\mathcal{K}$ stands for the kernel employed. For instance, when $\mathcal{K}(t)=\delta(t)$ the usual derivative is recovered. For the cases of the Caputo-Fabrizio and Atangana-Baleanu the kernel is given by
\begin{eqnarray}
\mathcal{K}_{CF}(t)=\frac{f(\alpha)}{1-\alpha} \exp \left[-\frac{\alpha t}{1-\alpha} \right],
\label{Kcaputo}
\end{eqnarray}
for $0<\alpha<1$ \cite{caputo2015new}, and
\begin{eqnarray}
\mathcal{K}_{AB}(t)=\frac{b(\alpha)}{1-\alpha}E_{\alpha}\left[-\frac{\alpha t^{\alpha}}{1-\alpha} \right],
\label{KAtangana}
\end{eqnarray}
for $0<\alpha<1$ with $E_{\alpha}(\cdots)$ the Mittag-Leffler function \cite{haubold2011mittag} which is the kernel associated to the fractional operator of Atangana and Baleanu \cite{atangana2016new}. The AB (Atangana--Baleanu) fractional derivative has two characteristic: a non--singular form (i.e. $\lim_{t \rightarrow 0}\mathcal{K}_{AB}(t) \neq +\infty$) and a power law behaviour for long times, i.e. $\mathcal{K}_{AB}(t) \propto t^{-1-\alpha}$.

\subsection{Fractional Fokker-Planck equation}\label{subsec2.2}

The progress in the fractional calculation was accompanying the developments obtained in the field of diffusion differential equations. Fractional derivatives offer multiples ways of generalising diffusion differential equations as well as for characterising more classes of phenomena than the standard diffusion equation. The unidimensional Fokker-Planck equation with constant diffusion coefficient $D_0$
\begin{eqnarray}\label{FPE}
\frac{\partial \rho(x,t)}{\partial t}=D_0 \frac{\partial^2\rho(x,t)}{\partial x^2}
\end{eqnarray}
describes the evolution of the probability density function (PDF) $\rho(x,t)$ of finding a particle subjected a drag and random forces. The Eq. (\ref{FPE})
describes normal diffusion and further generalisations are needed for taking into account other types of diffusion that can be occur. Even more, the temporal derivative
$\frac{\partial \rho(x,t)}{\partial t}$ is not appropriate to consider memory effects that are present
in a wide class of phenomena. Here is when the fractional generalisations of the FPE enter in order to characterise memory effects of the evolution of the PDF.
%that cannot be described and other types of diffusion than the normal or anomalous ones.
For accomplish this, we consider the following generalisation of the FPE (\ref{FPE}) (that we called fractional FPE)
\begin{eqnarray}
\mathcal{O}^{\mathcal{K}}_t\rho(x,t)= \frac{\partial}{\partial x}
\left(\sqrt{\mathcal{D}(x)} \frac{\partial }{\partial x} \left(\sqrt{\mathcal{D}(x)}\rho(x,t)\right)\right),
\label{diffusion}
\end{eqnarray}
where $\mathcal{O}^{\mathcal{K}}_t$  is the
fractional derivative with kernel $\mathcal{K}$ given by the Eq. (\ref{fractional-derivative-kernel})
and $\mathcal{D}(x)$ is a non-constant diffusion coefficient. It is clear that for $\mathcal{K}(t)=\delta(t)$
and $\mathcal{D}(x)=D_0$ the traditional FPE (\ref{FPE}) is recovered.

\section{Fractional FPE with
non-singular kernel operators
}\label{sec3}

In this Section we obtain the solutions of the fractional FPE (\ref{diffusion}) in the context of non-singular operators
with the initial condition $\rho(x,0)=\delta(x)$. The calculations are performed for an arbitrary diffusion coefficient. Then, we consider a power-law dependence of the diffusion coefficient for a detailed analysis. Applying the Laplace transform of Eq. (\ref{diffusion})
we obtain
\begin{eqnarray}
s\mathcal{K}(s)\rho(x,s)-
\mathcal{K}(s) = \frac{\partial}{\partial x}\sqrt{\mathcal{D}(x)}   \frac{\partial}{\partial x} \sqrt{\mathcal{D}(x)} \rho(x,s).
\label{diffusionlanplace}
\end{eqnarray}
where $\mathcal{L}\lbrace \mathcal{O}_t^{\mathcal{K}} f(t) \rbrace = s\mathcal{K}(s) f(s)-
\mathcal{K}(s)f(0)$ is the
Laplace transform of the fractional derivative with kernel $\mathcal{K}$ (Eq. \ref{fractional-derivative-kernel}).
The Laplace transform of a function $f(t)$ is defined by $\int_0^{\infty}dt e^{-st} f(t)=\mathcal{L}\lbrace f(t) \rbrace= F(s)$ and its inversion is given by $\frac{1}{2i \pi}\lim_{T\rightarrow \infty}\int_{\sigma-i T}^{\sigma + i T}ds e^{st} F(s)=\mathcal{L}^{-1}\lbrace F(s)\rbrace =f(t)$, in which $s$ is a complex variable and $\sigma$ is a real number so that the contour path of integration in the region of convergence of $F(s)$.
We search for solutions of the Eq. (\ref{diffusionlanplace}) in the form of the \textit{ansatz}
\begin{eqnarray}
\rho(x,s)= \frac{\Upsilon(s)}{\sqrt{\mathcal{D}(x)}} \exp \left[-\int_{0}^{|x|}dx' \frac{\sqrt{s \mathcal{K}(s)}}{\sqrt{\mathcal{D}(x')}} \right]
\label{ansatz}\end{eqnarray}
where $\rho(x,s)=\rho(-x,s)$ points out the symmetry in the variable $x$. Establishing the change of variable in $x$
\begin{eqnarray}
y(x)=\int_{0}^{|x|}dx' \frac{1}{\sqrt{\mathcal{D}(x)}}, \qquad \textnormal{in which } \qquad \lim_{x \rightarrow \infty}y(x)=\infty,
\end{eqnarray}
and using the \textit{ansatz} (\ref{ansatz}) in Eq. (\ref{diffusionlanplace}) (with the normalization condition $\int_{-\infty}^{\infty} dx \rho(x,t)=1$) we obtain
\begin{eqnarray}
\Upsilon(s)=\frac{1}{2}\sqrt{\frac{\mathcal{K}(s)}{s}},
\label{upslon}
\end{eqnarray}
which implies a general solution in the Laplace coordinate space $s$
\begin{eqnarray}
\rho(x,s)=\frac{\sqrt{\mathcal{K}(s)}}{2 \sqrt{s \mathcal{D}(x)}}  \exp \left[-\int_{0}^{|x|}dx' \frac{\sqrt{s \mathcal{K}(s)}}{\sqrt{\mathcal{D}(x')}}  \right].
\label{ansatzlaplace}
\end{eqnarray}
For the fractional Caputo operator, defined by (\ref{fractional-derivative-kernel}) with the kernel $\mathcal{K}_C(t)=\frac{t^{-\alpha}}{\Gamma[1-\alpha]}$, the Laplace inversion of (\ref{ansatzlaplace})
gives the solution
\begin{eqnarray}
\rho(x,t)=\frac{1}{2 t^{\frac{\alpha}{2}}\sqrt{\mathcal{D}(x)}} {\mbox{\Large{H}}}^{1,0}_{1,1}\left[ \int_{0}^{|x|}dx' \frac{1}{\sqrt{\mathcal{D}(x')}} \frac{1}{t^{\frac{\alpha}{2}}} \Big| ^{\left(1-\frac{\alpha}{2},\frac{\alpha}{2}\right)}_{(0,1)} \right],
\label{caputo}
\end{eqnarray}
where the $\textnormal{\large{H}} (\cdots)$ is the Fox function \cite{mainardi2005fox} that consists of a type of Merlin integration. The solution (\ref{caputo}) is a non-Gaussian distribution. For a constant diffusion coefficient it is a typical solution of the diffusion equation proposed in Montroll works for describing continuous random walks in time \cite{montroll1965random,scher1975anomalous}. On the other hand, the solution (\ref{caputo}) with $\alpha=1$ contains a variety of PDF describing diffusion processes in a heterogeneous medium, which was investigated by Metzler \textit{et al} in \cite{metzler2014anomalousd}. In this framework, our intention is extend this first result to non-singular diffusion equations. This is not a non-trivial issue because, as we have seen, it is necessary to employ the Laplace-inverse transform on the $s$-complex plane and this typically can be a complicated task.

\subsection{General solution for the Caputo-Fabrizio operator}
\label{subsec3.1}

\begin{figure}[h!]
%\centering
\includegraphics[scale=0.8]{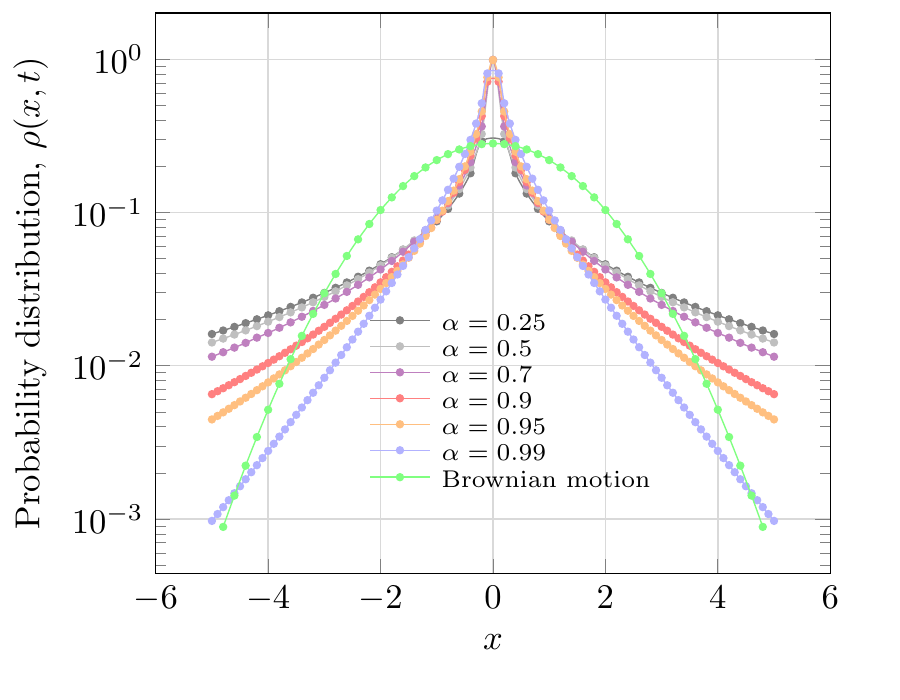}
\includegraphics[scale=0.8]{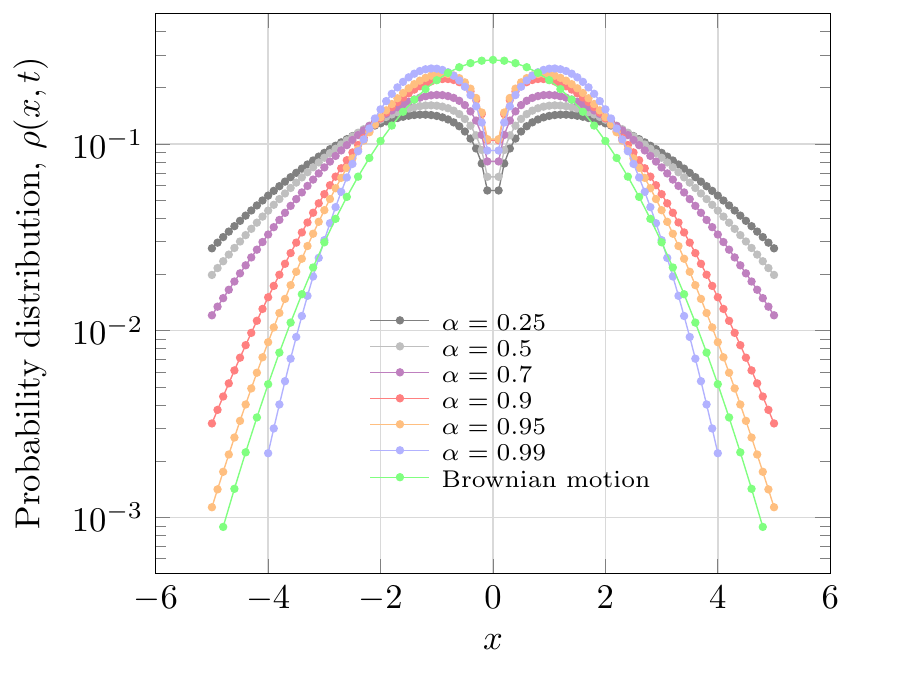}
\caption{Log of the solutions of the fractional FPE
for the Caputo-Fabrizio kernel (Eq. \ref{pdfcaputo}) at time $t=10^{-1}$ in function of the coordinate space $x$ for the values of the fractional index $\alpha=0.25,0.5,0.75,0.9,0.95,0.99$ and considering the parameter values $\frac{\mathcal{D}_0}{f(\alpha)} = 10$. For comparing, in both plots the Gaussian PDF (quadratic curve in green dots) with $D_0=10$ is illustrated. Left pannel: family of solutions for $\nu=1$ exhibiting a unimodal behavior. Right pannel: family of solutions for $\nu=-1$ exhibiting a bimodal behavior.}
\label{fig1}
\end{figure}

The kernel of the Caputo-Fabrizio fractional derivative can be written in the Laplace coordinate space $s$ as
\begin{eqnarray}
\mathcal{K}(s)=\frac{f(\alpha)}{1-\alpha}\frac{1}{s+\frac{\alpha}{1-\alpha}}.
\label{kernel1s}
\end{eqnarray}
which if replaced
in Eq. (\ref{ansatzlaplace}) allows us to obtain $\rho(x,s)$ in the compact form
\begin{eqnarray}
\rho(x,s)= -\frac{1}{2}\frac{\partial \Phi}{\partial |x|},
\label{pdfs}
\end{eqnarray}
where
\begin{eqnarray}
\Phi (x,s) = \frac{1}{s}\exp \left[ - \int_{0}^{|x|}dx' \frac{1}{\sqrt{\mathcal{D}(x')}} \sqrt{\frac{f(\alpha)}{1-\alpha}} \sqrt{\frac{s}{s+\frac{\alpha}{1-\alpha}}}\right].
\end{eqnarray}
Now, by calculating the inversion Laplace transform of this expression and taking the branch cuts associated with $\sqrt{s}$ and $\sqrt{s+\frac{\alpha}{1-\alpha}}$ along the negative real axis of $s$--plane for $\frac{\alpha}{1-\alpha} >0 $ (see Ref. \cite{widder2015laplace} for more details) we have
\begin{eqnarray}\label{phi-caputo}
\Phi (x,t) &=& 1- \frac{2}{\pi} \int_0^{\infty}d\eta \exp \left[-\frac{\alpha}{1-\alpha} \frac{t\eta ^2}{\eta ^2+1} \right] \frac{\mathcal{S}(t,\eta)}{\eta(1+\eta^2)},
\end{eqnarray}
with
\begin{eqnarray}\label{sin-caputo}
\mathcal{S}(t,\eta) &= & \sin \left[ \int_{0}^{|x|} dx' \frac{1}{\sqrt{\mathcal{D}(x')}} \sqrt{\frac{f(\alpha)}{1-\alpha}}  \eta \right].
\end{eqnarray}
Then, using Eqs. (\ref{phi-caputo}) and (\ref{sin-caputo}) in (\ref{pdfs}), we obtain the general solution of the fractional FPE (\ref{diffusion}) corresponding to the Caputo-Fabrizio kernel
\begin{eqnarray}
\rho(x,t) &=& \frac{1}{\pi}\sqrt{\frac{f(\alpha)}{(1-\alpha)\mathcal{D}(x)}} \int_0^{\infty}d\eta \exp \left[-\frac{\alpha}{1-\alpha} \frac{t\eta ^2}{\eta ^2+1} \right]\frac{\mathcal{C}(t,\eta) }{1+\eta^2},
\label{pdfcaputo}
\end{eqnarray}
with
\begin{eqnarray}
\mathcal{C}(t,\eta) &=& \cos \left[\eta \int_{0}^{|x|}dx' \frac{1}{\sqrt{\mathcal{D}(x')}} \sqrt{\frac{f(\alpha)}{1-\alpha}} \right].
\end{eqnarray}
This PDF implies a generalised non-Gaussian form for any choice of the fractional index $\alpha$, as we shall see.
Henceforward we consider the power-law form for the diffusion coefficient
\begin{eqnarray}
\mathcal{D}(x)=\mathcal{D}_0|x|^{\nu}, \qquad \nu<2.
\label{diffusioncoeficiente1}
\end{eqnarray}
that
was used by Andrey G. Cherstvy \textit{et. al} in a breaking ergodicity context \cite{cherstvy2013anomalous} and in other scenarios in \cite{sandev2017heterogeneous, hormuth2017mechanically,metzler2014anomalousd,lanoiselee2018model}.
Figs \ref{fig1} shows the logarithm of the PDF (\ref{pdfcaputo}) for the values
$\alpha=0.25,0.5,0.7,0.9,0.95,0.99$ and for $\nu=1$ and $\nu=-1$ at the left and right pannels respectively, along with the Gaussian PDF in green dots. We can see that both families of PDF ($\nu=1$ and $\nu=-1$) exhibit a non-Gaussian behavior with a long tail.
For $\nu=1$ every PDF presents an unimodal
shape while for $\nu=-1$ they present a bimodal one.

% \newpage
\subsection{General solution for the Atangana-Baleanu operator}
\label{subsec3.2}

\begin{figure}[h!]
\centering
\includegraphics[scale=0.8]{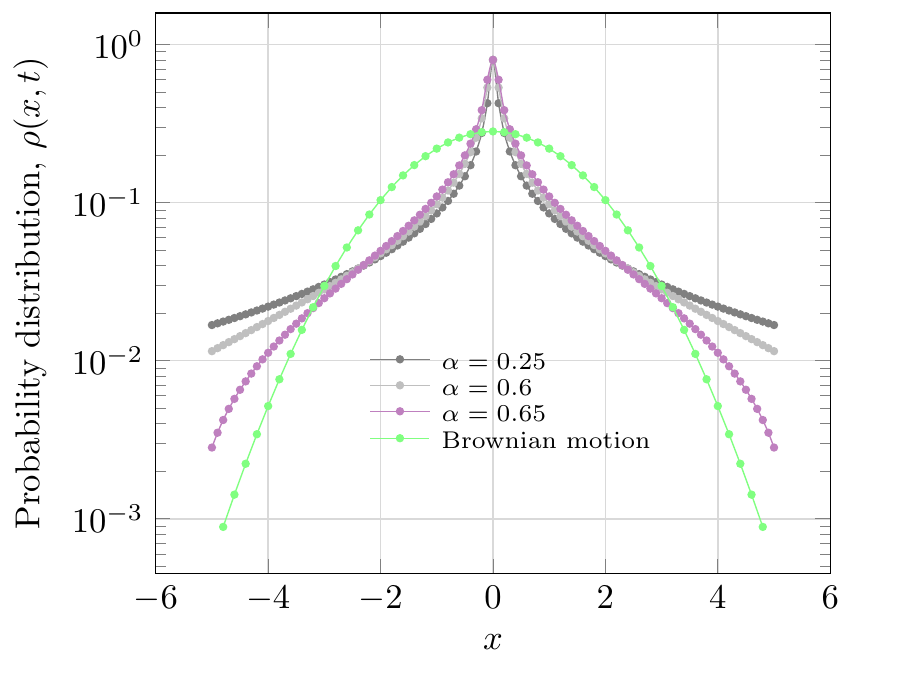}
\includegraphics[scale=0.8]{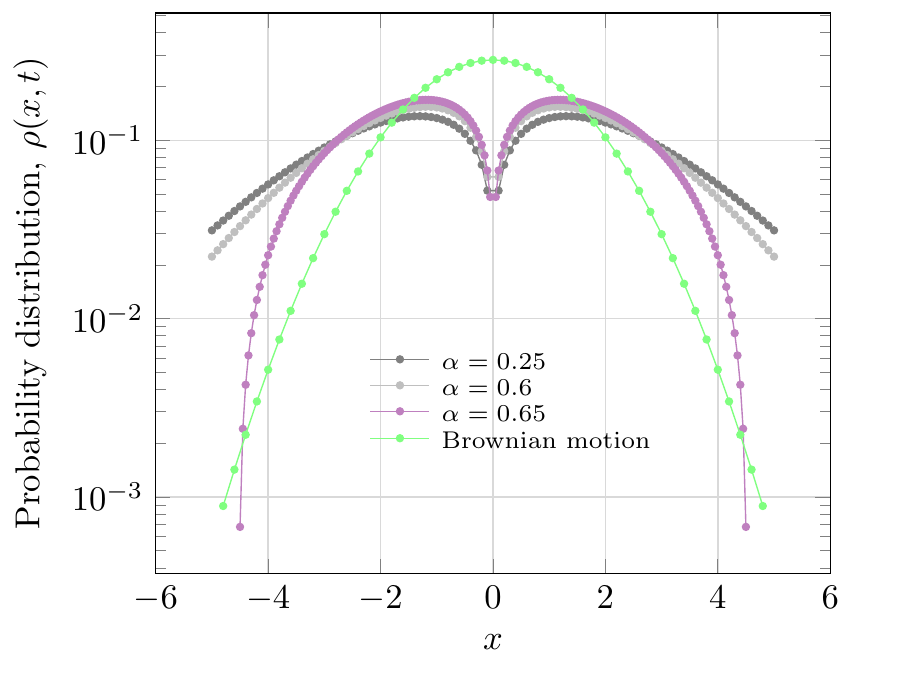}
\caption{Log of the solutions of the fractional FPE
for the Atangana-Baleanu kernel (Eq. \ref{pdfatagana}) at time $t=10^{-1}$ in function of the coordinate space $x$ for the values of the fractional index $\alpha=0.25,0.6,0.65$ and considering the parameter values $\frac{\mathcal{D}_0}{f(\alpha)} = 10$. For comparing, in both plots the Gaussian PDF (quadratic curve in green dots) with $D_0=10$ is illustrated. Left pannel: family of solutions for $\nu=1$ exhibiting a unimodal behavior. Right pannel: family of solutions for $\nu=-1$ exhibiting a bimodal behavior.}
\label{fig2}
\end{figure}

The kernel of the Atangana-Baleanu fractional derivative in the Laplace coordinate space $s$ is given by
\begin{eqnarray}\label{atangana-laplace-kernel}
\mathcal{K}(s)=\frac{b(\alpha)}{\alpha-1}\frac{s^{\alpha-1}}{s^{\alpha}+\frac{\alpha}{1-\alpha}},
\end{eqnarray}
which if replaced
in Eq. (\ref{ansatzlaplace}) allows us to obtain
$\rho(x,s)$, and then the solution $\rho(x,t)$ follows using the inverse Laplace transform.
Analogously to the Caputo-Fabrizio case,
$\rho(x,s)$ admits the representation expressed by the
Eq. (\ref{pdfs}) with $\Phi(x,s)$ given by
\begin{eqnarray}
\Phi (x,s) = \frac{1}{s}\exp \left[ - \int_0^{|x|}dx' \frac{1}{\sqrt{\mathcal{D}(x')}}  \sqrt{\frac{b(\alpha)}{1-\alpha}} \sqrt{\frac{s^{\alpha}}{s^{\alpha}+\frac{\alpha}{1-\alpha}}}\right],
\end{eqnarray}
which can be solved by integrating the branch cut associated with $s^{\alpha}$. Then, from the inverse Laplace transform we have
\begin{eqnarray}
\Phi(x,t) &=& 1-\frac{1}{\pi}\int_0^{\infty} e^{-t \eta}\exp \left \lbrace - \int_0^{|x|}dx' \frac{1}{\sqrt{\mathcal{D}(x')}} \sqrt{\frac{b(\alpha)\eta^{\alpha}}{\gamma_1}} \cos \left[ \frac{\alpha \pi - \gamma_2}{2} \right] \right \rbrace \nonumber \\
& \times &  \sin \left \lbrace \int_0^{|x|}dx' \frac{1}{\sqrt{\mathcal{D}(x')}}  \sqrt{\frac{b(\alpha)\eta^{\alpha}}{\gamma_1}} \sin \left[ \frac{\alpha \pi - \gamma_2}{2} \right] \right \rbrace \frac{d\eta}{\eta}
\label{up2}
\end{eqnarray}
with
\begin{eqnarray}
\gamma_1=\sqrt{\eta^{2\alpha}+\left(\frac{\alpha}{1-\alpha}\right)^2 + 2 \left(\frac{\alpha}{1-\alpha}\right) \eta^{\alpha} \cos[\alpha \pi]}
\end{eqnarray}
 and
\begin{eqnarray}
 \gamma_2= \arctan \left[\eta^{\alpha} \sin[\alpha \pi] \left(\eta^{\alpha}\cos[\alpha \pi]+\frac{\alpha}{1-\alpha}\right) \right].
 \end{eqnarray}
Using Eqs. (\ref{pdfs}) and (\ref{up2})
the general solution of the fractional FPE (\ref{diffusion}) corresponding to the Atangana-Baleanu kernel results
\begin{eqnarray}
\rho(x,t)&=&\frac{1}{2\pi}\int_0^{\infty} e^{-t \eta}\exp \left \lbrace -\int_0^{|x|}dx'\sqrt{\frac{b(\alpha)\eta^{\alpha}}{\mathcal{D}(x')\gamma_1}} \cos \left[ \frac{\alpha \pi - \gamma_2}{2} \right] \right \rbrace  \sqrt{\frac{b(\alpha)\eta^{\alpha}}{\mathcal{D}(x)\gamma_1}}  \nonumber \\
& \times & ( \cos \left[ \frac{\alpha \pi - \gamma_2}{2} \right]  \sin \left \lbrace \int_0^{|x|}dx' \sqrt{\frac{b(\alpha)\eta^{\alpha}}{\mathcal{D}(x')\gamma_1}} \sin \left[ \frac{\alpha \pi - \gamma_2}{2} \right] \right \rbrace \nonumber \\
& + &  \sin \left[ \frac{\alpha \pi - \gamma_2}{2} \right]   \cos \left \lbrace \int_0^{|x|}dx' \sqrt{\frac{b(\alpha)\eta^{\alpha}}{\mathcal{D}(x')\gamma_1}} \sin \left[ \frac{\alpha \pi - \gamma_2}{2} \right] \right \rbrace )   \frac{d\eta}{\eta},
\label{pdfatagana}
\end{eqnarray}
which is valid for all diffusion coefficient $\mathcal{D}(x)$. Assuming
$\mathcal{D}(x)=\mathcal{D}_0|x|^{\nu}$ with $\nu<2$ (as in Eq. (\ref{diffusioncoeficiente1})) and letting $\alpha \rightarrow 1$
the solution becomes in the exponential form
\begin{eqnarray}
\rho(x,t) = \frac{\exp\left[-\frac{\left| x\right| ^{2-\nu}}{ (2-\nu)^2 \mathcal{D}_0 t}\right]}{\sqrt{2 \pi\mathcal{D}_0 |x|^{\nu} t}},
\end{eqnarray}
which reproduces the standard Gaussian solution of the FPE for $\nu=0$.
Other representative case is obtained when
$\mathcal{D}(x)=constant$, which implies an analytical solution with an arbitrary fractional index $\alpha$
\cite{alkahtani2017generalized,hristov2017derivatives}.
Figs \ref{fig2} shows the logarithm of the PDF (\ref{pdfatagana}) for the values
$\alpha=0.25,0.6,0.65$ and for $\nu=1$ and $\nu=-1$ at the left and right pannels respectively along with the Gaussian PDF with green dots. Analogously as in the Caputo-Fabrizio case, we can see that both families of PDF ($\nu=1$ and $\nu=-1$) exhibit a non-Gaussian behavior with a long tail.
For $\nu=1$ every PDF presents an unimodal
shape while for $\nu=-1$ they present a bimodal one.

\section{Types of diffusion from fractional FPE: Caputo-Fabrizio and Atangana-Baleanu kernels}
\label{sec4}

Anomalous diffusion phenomena can be characterised by means of the MSD
of the particles, as was reported by Metzler and Klafter in \cite{metzler2000random}.
In our case, the general solution (\ref{ansatzlaplace}) implies
a null mean displacement $\langle x \rangle =0$, so MSD is equal to $\langle (x-\langle x \rangle)^2 \rangle =\langle x^2 \rangle $.
Using that the MSD is $\langle x^2 \rangle =\int_{-\infty}^{+\infty}x^2\rho(x,t)dx$
with the diffusion coefficient given by (\ref{diffusioncoeficiente1}), we obtain
for the general solution (\ref{ansatzlaplace}) that the MSD in the Laplace
space coordinate $s$ is
\begin{eqnarray}\label{anomalous-laplace}
\langle x^2 \rangle (s) = \frac{\sqrt{\mathcal{K}(s)} 2^{\frac{2 \nu}{\nu -2}} \Gamma \left[\frac{4}{2- \nu}\right] \left \lbrace \frac{\sqrt{\mathcal{K}(s) s}}{\sqrt{\mathcal{D}_0} (2- \nu)}\right \rbrace^{\frac{6-\nu }{\nu -2}}}{(2-\nu)^2 \sqrt{\mathcal{D}_0 s}}.
\end{eqnarray}
%To particular case in which $\nu=0$, we obtain $\langle x^2 \rangle (s) = \frac{2\mathcal{D}_0}{s^2\mathcal{K}(s)}$.
Considering the Caputo-Fabrizio kernel in the Laplace space coordinate $s$ (see Eq.(\ref{kernel1s})) and from the inversion Laplace transform of (\ref{anomalous-laplace}) we have
\begin{eqnarray}
\langle x^2 \rangle(t) =\int_0^{t}dt' \frac{e^{-\frac{\alpha  \left(t-t'\right)}{1-\alpha }} \left(\frac{t-t'}{1-\alpha }\right)^{-\frac{2}{2-\nu }-1}}{(1-\alpha ) \Gamma \left(\frac{2}{2-\nu }\right)} \frac{ 2^{\frac{2 \nu}{\nu -2}} \Gamma \left[\frac{4}{2- \nu}\right] t'^{\frac{2  }{2- \nu }} }{\Gamma\left[ \frac{4-\nu}{2-\nu} \right](2-\nu )^{\frac{\nu +2}{2-\nu }} \mathcal{D}_0^{\frac{2}{\nu-2}}},
\end{eqnarray}
that in the asymptotic limit of long times $t\rightarrow\infty$ becomes $\langle x^2 \rangle \sim t^{\frac{2 }{2- \nu }}$. This means that the fractional index of the Caputo-Fabrizio operator does not influence on the type of diffusion in the asymptotic limit, which is a similar behaviour present in clusters of percolation \cite{arkhincheev1991anomalous}.

Regarding the Atangana-Baleanu operator, whose kernel in the Laplace space coordinate $s$ is given by (\ref{atangana-laplace-kernel}), by introducing this in (\ref{anomalous-laplace}) and from its inversion Laplace transform we have
\begin{eqnarray}
\langle x^2 \rangle(t) \sim \left(\frac{1}{\alpha }\right)^{\frac{2}{\nu -2}} \frac{ 2^{\frac{2 \nu}{\nu -2}} \Gamma \left[\frac{4}{2- \nu}\right] t^{\frac{2 \alpha }{2- \nu }} }{\Gamma\left[ 1+\frac{2 \alpha }{2-\nu}  \right](2-\nu )^{\frac{\nu +2}{2-\nu }} \mathcal{D}_0^{\frac{2}{\nu-2}}},
\label{MSDAb}
\end{eqnarray}
where the normal diffusion is recovered for the combination $\alpha + \frac{\nu}{2}=1$.
%, which admits non-Gaussian forms.
Hence, from (\ref{MSDAb}) we obtain several types of diffusion
%implies in a all anomalous diffusive process
in function of the pair of parameters $(\alpha,\nu)$
\begin{eqnarray}\label{extensionMSD}
\lim_{t\rightarrow +\infty} \langle x^2 \rangle (t) \propto \left \{ \begin{array}{llll}  \mbox{subdiffusion}, & 2\alpha < 2-\nu \\ \mbox{superdiffusion}, & 4-2\nu > 2\alpha > 2-\nu
\\ \mbox{ballistic diffusion} , & \alpha = 2-\nu
\\ \mbox{hyperdiffusion} , & \alpha > 2-\nu.
\end{array} \right.
\end{eqnarray}
Interestingly, for a constant diffusion coefficient ($\nu=0$) we still have a range of
fractional indexes $\alpha$ where the  subdiffusion, superdiffusion, ballistic and hyper diffusion regimes can occur. This marks the degree of generality of the Arangana-Baleanu fractional operator.
Eq. (\ref{extensionMSD}) represents the class of diffusive processes associated to the non-singular Atangana--Baleanu fractional operator, typically applied on the irregular movement of Brownian particles in heterogeneous media.

\section{Conclusion}
\label{sec5}

In this work we have investigated the properties of the solutions of the fractional Fokker-Planck equation with non-singular fractional operators. Our study has been based on two non-singular types of memory kernels: the Caputo-Fabrizio operator and the Atangana-Baleanu one.
In both cases, with the help of the Laplace transform we have obtained the general and analytical solution
of the fractional FPE for an arbitrary diffusion coefficient $\mathcal{D}(x)$.
To illustrate the types of diffusion that can be characterised, we have considered a diffusion coefficient
of the power law type $\mathcal{D}(x)=\mathcal{D}_0|x|^{\nu}$.%

The solutions of the Caputo-Fabrizio operator resulted non-Gaussian distributions
that are strongly connected with the index $\alpha$ of the memory kernel,
having a unimodal or bimodal behavior if the diffusion index $\nu$ is positive or negative
respectively (see Fig. \ref{fig1}). However, in the asymptotic limit of long times
$t\rightarrow\infty$ the fractional index $\alpha$ does not affect
the type of diffusion involved.

For the case of the Atangana-Baleanu operator, their solutions
 also resulted non-Gaussian distributions
presenting a unimodal or bimodal behavior according if
the index $\nu$ is positive or negative
respectively (see Fig. \ref{fig2}).
Nevertheless, in the asymptotic limit $t\rightarrow\infty$ the type of diffusion has
a dependence on the pair of indexes $(\alpha,\nu)$ that allows a
characterisation of the normal diffusion on the curve $\alpha+\frac{\nu}{2}=1$ as
 well as the existence of sub-super diffusion and ballistic-hyper diffusion even
 for a constant diffusion coefficient ($\nu=0$).

We consider the results and techniques employed in this work constitute important tools for studying non-Markovian diffusive process with memory effects, thus opening new possibilities in futures researches concerning: diffusion to intermittent movement, reaction-diffusion systems and comb (or grid) structure under influence of new fractional derivative forms.

\section*{Acknowledgements}
M. A. F. dos Santos acknowledges the support of the Brazilian agency CNPq.
I. S. Gomez acknowledges fellowships received from CAPES / INCT-SC (at Universidade
Federal da Bahia).

\section*{References}
%\bibliographystyle{iopart-num}
%\bibliography{fractional-FPE}

\end{document}